# Simulations and experiments of short intense envelope solitons of surface water waves

A. Slunyaev[1,2], G.F. Clauss[3], M. Klein[3], M. Onorato[4]

[1] *Institute of Applied Physics, Nizhny Novgorod, Russia, Slunyaev@hydro.appl.sci-nnov.ru*

[2] *N. Novgorod Technical State University, Nizhny Novgorod, Russia*

[3] *Ocean Engineering Division, Technical University of Berlin, Germany, Klein@naoe.tu-berlin.de*

[4] *Universita di Torino, Dipartimento di Fisica Italy and Istituto Nazionale di Fisica Nucleare, Sezione di Torino, Miguel.Onorato@gmail.com*

The problem of existence of stable nonlinear groups of gravity waves in deep water is revised by means of laboratory and numerical simulations with the focus on intense waves. Wave groups with steepness up to $A_{cr}\, \omega_m^2/g \approx 0.30$ are reproduced in laboratory experiments ($A_{cr}$ is the wave crest amplitude, $\omega_m$ is the mean angular frequency and $g$ is the gravity acceleration). We show that the groups remain stable and exhibit neither noticeable radiation nor structural transformation for more than 60 wave lengths or about 15-30 group lengths. These solitary wave patterns differ from the conventional envelope solitons, as only a few individual waves are contained in the group. Very good agreement is obtained between the laboratory results and strongly nonlinear numerical simulations of the potential Euler equations. The envelope soliton solution of the nonlinear Schrödinger equation is shown to be a reasonable first approximation for specifying the wavemaker driving signal. The short intense envelope solitons possess vertical asymmetry similar to regular Stokes waves with the same frequency and crest amplitude. Nonlinearity is found to have remarkably stronger effect on the speed of envelope solitons in comparison to the nonlinear correction to the Stokes wave velocity.

**Keywords:** steep and short solitary wave groups, laboratory experiment, strongly nonlinear numerical simulations

**I. INTRODUCTION**

The paper is aimed at revising the conventional notion of envelope solitons in the context of intense surface water gravity wave groups. It is notorious that the envelope soliton is an exact solution of the integrable nonlinear Schrödinger equation (NLS), which is the basic model that describes weakly nonlinear water waves propagating in one direction [1,2]. Envelope solitons compose the major part of the wave field for the Cauchy problem in the NLS equation framework in the limit of very long time [3]. The possibility of solution of the NLS equation by means of the Inverse Scattering Transform makes envelope solitons extremely attractive objects which are useful for comprehension of nonlinear wave dynamics, see e.g. Refs. 4 and 5.

The modulational or Benjamin – Feir instability, known from 1960-s, is an efficient mechanism of water wave grouping under appropriate conditions (see e.g. review by Zakharov and Ostrovsky [6]). These groups are commonly observed in time series of sea surface elevation and give an idea how envelope solitons may become apparent in natural conditions.

Solitary groups from small to moderate amplitudes were tested in laboratory conditions, see Refs. 7, 8, 9; they were also examined in numerical simulations within different frameworks [10, 11, 12]. The general conclusion to be drawn from the studies is that when transverse direction effects are neglected, weakly nonlinear wave groups do exhibit some structural stability, they behave similar to as prescribed by the weakly nonlinear framework and propagate without noticeable distortion and survive after collisions. Simultaneously, relatively steep solitary groups, $k_0 A_0 >\sim 0.2$ (where $k_0$ is the carrier wave number, and $A_0$ is the soliton amplitude) disperse significantly, and thus loose energy.

Despite the preceding knowledge, very short wave groups which contain very steep waves (up to the breaking limit), have been found in recent fully nonlinear simulations of potential Euler equations by Dyachenko and Zakharov [13]. In Ref. 14 the short intense solitary wave groups were associated with the strongly nonlinear limit of envelope soliton solutions of the NLS equation. It was found that this analytic solution provides a rather accurate initial condition for simulating strongly nonlinear coherent wave groups, which behave similar to envelope solitons, even when the wave steepness approaches the value of about $k_0 A_0 \sim 0.3$. Third order bound waves determined within the generalized NLS theory (see e.g. Refs. 15-17), were employed to improve description of water waves. For steeper initial conditions the wave group underwent breaking at the early stage of evolution. Comparing the fully and weakly nonlinear simulations, it was found in Ref. 14 that the dynamics of these strongly nonlinear solitary wave groups was rather well captured by the modified NLS equation (the generalized Dysthe model [18, 15, 17]) up to the steepness about $k_0 A_0 \sim 0.2$.

Concluding, the question of existence of steep and short solitary wave trains has not been clear by present. On the one hand, recent fully nonlinear simulations of unidirectional waves predict existence of very steep and short solitary structures with the maximum amplitude which is limited due to the wave breaking effect. The results of numerical simulations allow to consider the NLS soliton solution as a reasonable first approximation for these groups even in the case of rather steep waves ($k_0 A_0 \sim< 0.2$). On the other hand, previous laboratory measurements of envelope solitons of water waves cannot confirm that steep stationary solitary groups do exist in nature.

The laboratory experiments reported in this paper pursue two main objectives:

- to reproduce the 'limiting' envelope solitons, i.e., very short groups of very intense waves (as intense as possible, with steepness exceeding the value of $k_0A_0 = 0.2$), which propagate with persistence;
- to test the NLS envelope soliton solution as the boundary condition for generating nonlinear solitary wave groups with the focus on the case of large wave steepness.

The steep solitary wave groups, which we consider, contain very small number of individual waves. Strongly nonlinear numerical simulations are employed in the present study with the primary intension to compare the laboratory results with the theory. Preliminary results of this study were reported in Ref. 19.

The paper is organized as follows. Section II reminds briefly some features of the weakly nonlinear theory described by the nonlinear Schrödinger (NLS) equation, in particular, the envelope soliton solution is introduced. Stationary nonlinear wave groups, which are obtained in numerical simulations of potential Euler equations, are analyzed and discussed in Section III. The experimental setup and results of laboratory measurements of intense wave groups are given in Section IV. They are also compared with the results of numerical simulations given in Section III. The Conclusion contains the summary of findings, and the discussion.

## II. WEAKLY NONLINEAR THEORY FOR ENVELOPE SOLITONS

The nonlinear Schrödinger (NLS) equation,

$$i\left(\frac{\partial A}{\partial t} + C_{gr}\frac{\partial A}{\partial x}\right) + \frac{\omega_0}{8k_0^2}\frac{\partial^2 A}{\partial x^2} + \frac{\omega_0 k_0^2}{2}|A|^2 A = 0, \qquad (1)$$

represents the leading-order theory for description of unidirectional gravity water waves under the assumptions of small nonlinearity, $k_0|A| \ll 1$ (where $A$ is the amplitude of water displacement), and narrow spectral bandwidth, $\Delta k /k_0 \ll 1$. Here the carrier wave number, $k_0$, and characteristic wavenumber spectrum width, $\Delta k$, are introduced. The group velocity of linear waves, $C_{gr} \equiv \omega_0 /(2k_0)$, is provided by the deep-water dispersion relation,

$$\omega_0 = \sqrt{gk_0} \, . \qquad (2)$$

In order to compare solutions of the weakly nonlinear theory with the results of laboratory experiments, it is useful to re-write the NLS equation (1) in the form, which describes evolution in space,

$$i\left(\frac{\partial A}{\partial x} + \frac{1}{C_{gr}}\frac{\partial A}{\partial t}\right) + \frac{\omega_0}{8k_0^2 C_{gr}^3}\frac{\partial^2 A}{\partial t^2} + \frac{\omega_0 k_0^2}{2C_{gr}}|A|^2 A = 0 \, . \qquad (3)$$

From mathematical point of view, equation (3) has the same domain of validity as (1).

The complex-valued function of coordinate and time, $A(x, t)$, determines both, the surface elevation, $\eta(x, t)$, and the velocity potential $\varphi(x, z, t)$. For a given function of time $A(t)$ or space $A(x)$ the fields $\eta$ and $\varphi$ may be computed with accuracy up to the third order of the asymptotic theory (i.e., including terms of order $O((k_0 A)^3)$ and $O((\Delta k /k_0)^3)$ by virtue of a little bit bulky but still straightforward reconstruction formulas, which may be found in Refs. 16, 17.

The classical NLS equation is able to describe nonlinear wave dynamics accurately for a rather short distance of wave propagation (a few times of $\sim A^{-2} k_0^{-3}$, according to the laboratory investigation in Ref. 20, what is of order of one characteristic 'nonlinear distance'). The Dysthe equation, which is the next-order generalization of the NLS model, provides much better description, see. e.g. Ref. 20, 21. Other generalizations of the high order NLS theory are possible; often they represent a reasonable compromise of efficiency and simplicity, and therefore are quite popular.

The NLS equation is integrable by means of the Inverse Scattering Technique and possesses the envelope soliton solution in the form of a sequence of waves which constitute a stable wave group (see e.g. Refs. 4, 5). For the equation in form (1) the envelope soliton solution is given by the expression

$$A(x,t) = A_0 \frac{\exp\left[i\frac{s_0^2}{4}\omega_0 t\right]}{\cosh\left[\sqrt{2}s_0 k_0 x\right]}, \quad (4)$$

where $A_0$ is the soliton amplitude; the characteristic wave steepness of the soliton may be specified by $s_0 \equiv k_0 A_0$. Strictly speaking, the NLS equation allows envelope solitons which may propagate with velocity different from the carrier wave speed. However, the best fit of the approximate linear dispersion dependence, prescribed by the NLS equation, to the primitive water wave equations is when the envelope soliton has the speed of the carrier wave, $C_{gr} = \omega_0/(2k_0)$, what is already implied by solution (4).

Within the framework of the integrable NLS equation envelope solitons interact elastically between each other and with other quasi-linear waves. In contrast to transient wave groups, the envelope soliton is constituted by interacting coherent wave harmonics, what prevents dispersion of the group. The Fourier spectrum of group (4) may be obtained straightforwardly,

$$\hat{A}(k,t) \equiv \int_{-\infty}^{\infty} A(x,t) e^{ikx} dx = B \exp(i\theta),$$

$$B(k) \frac{\pi A_0}{\sqrt{2}s_0 k_0} \frac{1}{\cosh\dfrac{\pi k}{2\sqrt{2}s_0 k_0}}, \qquad \theta(t) = \frac{s_0^2}{4}\omega_0 t. \quad (5)$$

Hence, all Fourier modes are always co-phased. It is remarkable, that the Fourier amplitudes $B(k)$ do not evolve in time, thus, the NLS envelope soliton corresponds to a stationary solution in the Fourier space with co-phased harmonics.

Because equation (3) differs from equation (1) only in coefficients, solutions (4) and (5) may by straightforwardly re-written for the boundary problem, described by (3).

## III. NUMERICAL SIMULATIONS

The strongly nonlinear algorithm for solving the potential Euler equations for infinitively deep water in a periodic spatial domain, the High Order Spectral Method (HOSM) in the treatment of West *et al* [11], is employed in the study. The nonlinear parameter, which controls the number of terms in the expansion of the velocity potential near the rest level, is put equal to six, $M = 6$, what corresponds to accounting for up to 7-wave interactions.

The envelope soliton (4) provides the initial condition for the numerical simulations. Namely, the surface elevation, $\eta(x)$ and the surface velocity potential, $\Phi(x) = \varphi(x, z = \eta)$, are used to start the simulation at $t = 0$; they are obtained when the three order bound waves are taken into account (which correspond to the second and third harmonic and also to the induced long-scale component of the surface displacement and velocity potential [16, 17]). Value of the carrier wavenumber $k_0 = 1$ rad/m is used in all numerical simulations; the gravity acceleration is specified as $g = 9.81$ ms$^{-2}$.

Intense solitary groups were simulated in Ref. 14 within the framework of the HOSM and also using a fully nonlinear code of the Euler equations in conformal variables. Both the approaches showed quite similar wave dynamics, i.e., at the early stage of the evolution the initial wave group underwent some relatively weak radiation in both directions (quasi linear waves which propagate with group velocities both larger and smaller than that of the soliton); later on the radiated component was spread throughout the computational domain. If the initial wave group was steeper than about $k_0 A_0 \approx 0.3$, then the wave breaking was observed shortly after the start of simulations. If the wave survived after the initial stage, then the solitary wave group got stabilized and was propagating over the background of the radiated component for a long time without any evidence of energy leakage. As the radiated component observed in Ref. 14 was reasonably small, the NLS solution is considered hereby as the first approximation for steep solitary wave groups. In contrast to Ref. 14, in the present study a moving padding frame is introduced, which damps all waves aside some area close to the wanted wave group. A qualitatively similar approach was used by Dyachenko and Zakharov [13].

The evolution of the wave group is first simulated for a couple hundred of wave periods with the purpose to let it adjust and to obtain the shape of the 'true' stationary wave group. During this period some wave energy which is radiated away from the wanted group is adsorbed due to the padding mask, and hence the total wave energy decays with time, see solid line in Figure 1. In this figure temporal evolutions of the potential (dash-dotted line), kinetic (broken line) and total (solid line) energies are shown normalized by the corresponding initial values.

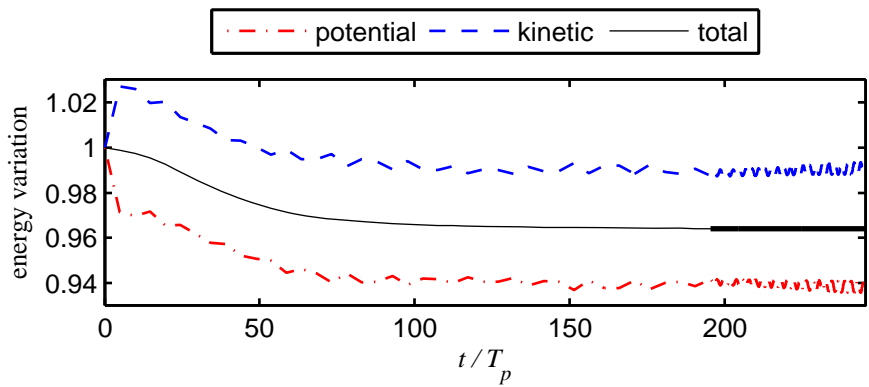

FIG. 1. Evolution of the potential, kinetic and total energies (see the legend) in numerical simulations; variations with respect to the corresponding initial values. Experiment No 9 from Table I is shown ($k_0 A_0 = 0.3$).

The figure reveals that the initial condition does not provide the proper balance between potential and kinetic energies, as the corresponding curves immediately get away from each other and preserve the difference. The total wave energy loss in the case shown in Figure 1 is about 3%, and is about twice smaller than the difference between the kinetic and potential energies. Figure 1 reports on the case of fairly intense waves, $k_0A_0 = 0.3$; other properties of the observed stationary wave group are given in Table I, see experiment No 9. The energy dissipation and the difference between kinetic and potential energies are smaller if less intense waves are simulated; the latter confirms that nonlinear effects are the reason for the observed difference between the potential and kinetic parts of the energy.

After the stage of adjustment (for about 200 wave periods), we start to store the simulated data with short time intervals (for the following about 50 wave periods) with the purpose to obtain the high-resolution picture of the wave dynamics. For this stage the total energy is shown by a thicker line in Figure 1. Some small fluctuations of potential and kinetic energies within the scale of a wave period may be seen in the figure. The values of the energies are almost constant for $t > 200\ T_p$, which implies an almost stationary state. In all cases solitary wave groups were formed in the course of the evolution. The data accumulated during the stage from 200 to 250 wave periods is used for the investigation of properties of the stationary wave groups.

Rather steep wave groups are simulated in the paper, with the initial condition in form of a NLS soliton (4) and characteristic steepness in range $0.15 \leq k_0A_0 \leq 0.35$, see Table I. Instability of numerical simulations, associated with the wave breaking phenomenon, is observed in the case $k_0A_0 = 0.35$. The steepest reliably stable case in the simulations corresponds to $k_0A_0 = 0.32$. Hence, some natural maximum amplitude limit of the solitary wave groups is found in the interval of initial parameters $0.32 < k_0A_0 < 0.35$.

The evolution of different characteristics of wave intensity is displayed in Figures 2a,b for cases $k_0A_0 = 0.2$ and $k_0A_0 = 0.3$ respectively. Four characteristics of the wave magnitude are considered: i) wave crest steepness, $k_mA_{cr}$, ii) wave trough steepness, $k_mA_{tr}$,

iii) steepness specified on the basis of the dimensionless half-height, $k_m H_x /2$, and the maximum local slope of the surface displacement, $\partial \eta / \partial x$. Quantities $A_{cr}$ and $A_{tr}$ are the zero-crossing crest and trough amplitudes respectively. Hereafter, $H_x$ and $H_t$ correspond to the wave heights determined on the basis of spatial or time series respectively. Each of them corresponds to the maximum value between the up-crossing and down-crossing heights. The frequency and wavenumber Fourier spectra of the stationary wave groups are used to obtain the peak ($\omega_p$, $k_p$) and mean ($\omega_m$, $k_m$) values of the cyclic frequency and the wavenumber respectively. The mean values $\omega_m$ and $k_m$ are obtained as the first moments of the corresponding power spectrum, and are found to be more consistent than the peak values, see Table I. These quantities differ from the initially specified carrier wavenumber, $k_0$, and the corresponding frequency of linear waves, specified by (2). In the simulations values $\omega_p$ are found to be slightly smaller than the linear frequency of initial waves, $\omega_0$. The wave number $k_p$ coincides with $k_0$ at $t = 0$, but later on the actual peak wavenumber, $k_p$, becomes slightly less than $k_0$. The mean frequency and wavenumber are slightly larger than the peak values, see Table I. The corresponding wave periods are defined by $T_p = 2\pi/\omega_p$ and $T_m = 2\pi/\omega_m$; the peak wave length is $\lambda_p = 2\pi/k_p$.

Amplitude characteristics of the steady nonlinear groups observed in numerical simulations are shown in Figure 2 as functions of time. It may be pointed out that these characteristics attain local extremes at close but different times; the wave height oscillates twice more frequent than other characteristics. Some weak variability of the wave group during the displayed time interval is quite obvious in Figure 2b.

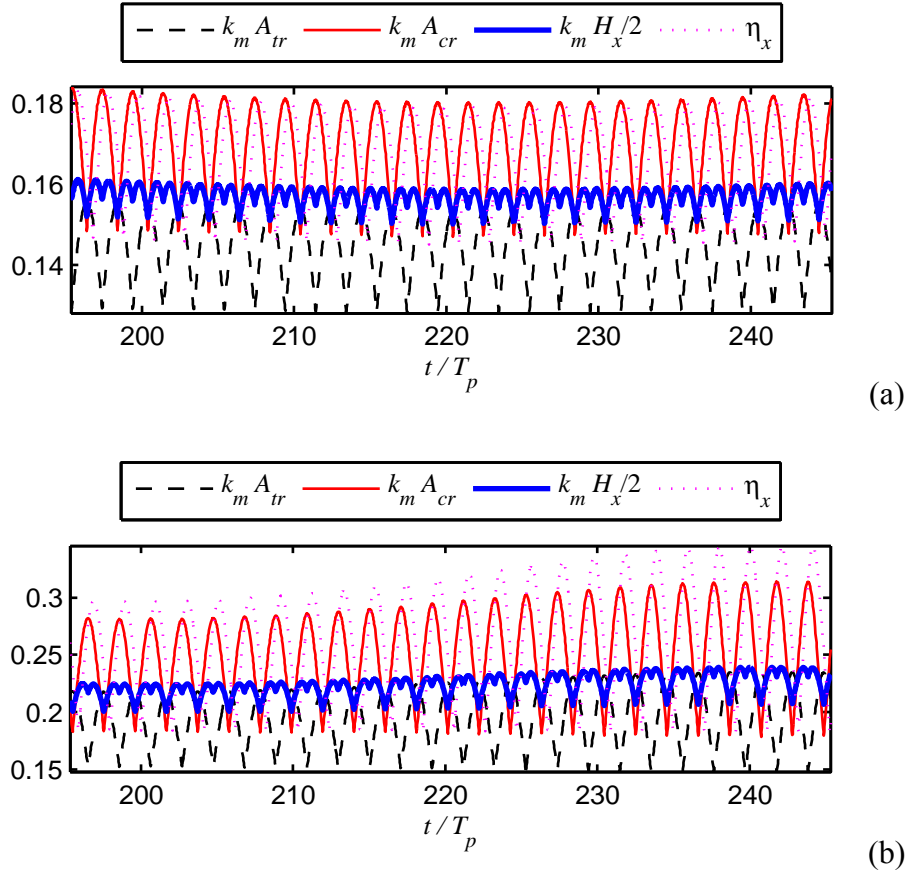

FIG. 2. Evolution of characteristics of the stationary wave group intensity as function of time. The steepness of wave troughs and wave crests are shown by dashed and thin solid lines correspondingly; estimation of the steepness on the basis of the wave height is given by thick solid line; dots show maximum local slope of the surface displacement. Cases $k_0 A_0 = 0.2$ (a) and $k_0 A_0 = 0.3$ (b) (experiments No 3 and 9 from Table I respectively).

The description of surface displacement by the nonlinear Schrodinger equation may be improved when phase-locked bound wave components are taken into account, see e.g., Refs. 16, 17. For simplicity, in estimations below we take into account bound waves for uniform Stokes waves up to the cubic nonlinear correction. The bound wave components alter the Stokes wave shape as follows,

$$kA_{cr}^{(3)} = kA + \frac{1}{2}k^2A^2 + \frac{3}{8}k^3A^3, \qquad kA_{tr}^{(3)} = kA - \frac{1}{2}k^2A^2 + \frac{3}{8}k^3A^3, \qquad (6)$$

where $k$ is the characteristic wave number.

For the case $k_0A_0 = 0.3$ (see experiment No 9 from Table I) the initial wavenumber is $k_0 = 1$ rad/m, and the wavenumber of the stationary wave group in the simulations is $k_p \approx 0.92$ rad/m, and therefore $k_pA_0 \approx 0.276$. The steepness of the maximum wave in the initial group is estimated with use of (6) as $k_0A^{(3)}_{cr} \approx 0.355$ and $k_0A^{(3)}_{tr} \approx 0.265$, and for the actual peak wavenumber of the solitary group the values are $k_pA^{(3)}_{cr} \approx 0.323$ and $k_pA^{(3)}_{tr} \approx 0.246$. Hence, Figure 2b reports on significant decay of the wave steepness in comparison to the initial condition.

A similar estimation for the case $k_0A_0 = 0.2$ (experiment No 3 from Table I) gives values $k_0A^{(3)}_{cr} \approx 0.223$, $k_0A^{(3)}_{tr} \approx 0.183$ versus $k_pA^{(3)}_{cr} \approx 0.215$ and $k_pA^{(3)}_{tr} \approx 0.178$. The values for other numerical experiments are given in Table I.

Due to the strong asymmetry of waves, it is not straightforward to apply the concept of wave envelope. In the present study wave envelopes are obtained by assembling all the surface elevations during the last 50 wave periods of simulations and all coordinates $x$, plotted in the reference frame, which moves with velocity of the group, $V$. This velocity is obtained as the speed of propagation of the wave group 'center of mass',

$$x_s(t) = \frac{\int x\eta^2(x,t)dx}{\int \eta^2(x,t)dx}, \qquad (7)$$

where the integration is performed within the spatial domain of the computation, taking into account the periodic boundary conditions. The sequence of positions $x_s(t)$ is interpolated by the relation $x_s = x_0 + Vt$, what gives the wanted value of the wave group velocity, $V$. The obtained solitary group velocities (see Table I) are discussed in the next Section when compared with results of the laboratory measurements (see Figure 9).

The observed stationary groups are very short wave packets of very intense waves, and hence the concepts of a wave modulation and a wave envelope cannot be applied to the case straightforwardly. The maxima and minima of the surface elevations for a given coordinate/time in the co-moving frame provide the upper and lower enveloping curves correspondingly. These envelope shapes for stationary wave groups as functions of time are shown in Figure 3a. The stationary wave groups shown in Figure 3a result from initial conditions in the form of NLS solitons with parameters $k_0 A_0$ from 0.15 to 0.32, see Table I. The envelopes become higher and shorter as the initial steepness is larger.

Figure 3b estimates the vertical asymmetry of the envelopes (stars). Note that here values $A_{cr}$ and $A_{tr}$ are amplitudes of the upper enveloping curve in Figure 3a, and of the corresponding lower one. In other words, values $A_{cr}$ and $A_{tr}$ are the maximum crest height and the deepest trough of the individual waves which may be observed in the envelope in the course of its evolution. The horizontal axis in Figure 3b measures dimensionless crest amplitudes, which are scaled with the help of the peak frequency. These characteristics of wave shapes are collected in Table I, but scaled with the help of the mean frequency. Quantity $\max(H_t)$ estimates the maximum dimensionless wave height in time series of wave groups.

The vertical asymmetry of uniform Stokes waves, $A_{cr} / A_{tr}$, is given by the dashed line in Figure 3b for the reference; for this line the horizontal axis measures the wave crest amplitude, scaled in a similar fashion, $A_{cr}\omega_p^2/g$, where $\omega_p$ is the frequency of the Stokes wave with the nonlinear correction taken into account (the strongly nonlinear correction to the Stokes wave frequency is found numerically). When the horizontal axis in Figure 3b uses the scale of $\omega_m$ instead of $\omega_p$, the symbols are placed a little bit lower than the dashed line. Estimations of the mean frequency and the mean wavenumber are assumed to be more reliable than the ones of the peak values. They provide with more definite conclusions on the comparison between the results of numerical and laboratory experiments. These mean characteristics are used in the next Section, where laboratory measurements of propagating intense wave groups are discussed.

Figure 3b claims that the vertical asymmetry of the solitary wave groups very well agrees with the asymmetry of Stokes waves with the same crest amplitudes and the same peak frequencies. In accordance with this statement, the envelope upper amplitude, $A_{cr}$, will be used as a measure of wave group intensity, which is consistent with the Stokes wave crest amplitude, $A_{cr}$.

Table I. Characteristics of stationary wave groups observed in numerical simulations.

| Numerical experiment No | $k_0A_0$ | $k_0A^{(3)}_{cr}$ | $k_0A^{(3)}_{tr}$ | $A_{cr}\dfrac{\omega_m^2}{g}$ | $A_{tr}\dfrac{\omega_m^2}{g}$ | $\max(H_t)\dfrac{\omega_m^2}{2g}$ | $k_p$, rad/m | $k_m$, rad/m | $\omega_p$, rad/s | $\omega_m$, rad/s | $V$, m/s |
|---|---|---|---|---|---|---|---|---|---|---|---|
| 1 | 0.15 | 0.16 | 0.14 | 0.15 | 0.13 | 0.14 | 0.99 | 0.99 | 3.13 | 3.16 | 1.60 |
| 2 | 0.16 | 0.17 | 0.15 | 0.16 | 0.14 | 0.15 | 1.01 | 0.99 | 3.13 | 3.16 | 1.61 |
| 3 | 0.20 | 0.22 | 0.18 | 0.19 | 0.16 | 0.17 | 0.97 | 0.98 | 3.07 | 3.17 | 1.62 |
| 4 | 0.22 | 0.25 | 0.20 | 0.21 | 0.18 | 0.20 | 0.97 | 0.98 | 3.13 | 3.18 | 1.63 |
| 5 | 0.23 | 0.26 | 0.21 | 0.23 | 0.19 | 0.21 | 0.97 | 0.98 | 3.13 | 3.20 | 1.64 |
| 6 | 0.25 | 0.29 | 0.22 | 0.27 | 0.21 | 0.23 | 0.88 | 0.98 | 3.19 | 3.22 | 1.66 |
| 7 | 0.28 | 0.33 | 0.25 | 0.28 | 0.22 | 0.26 | 0.98 | 0.98 | 3.07 | 3.22 | 1.67 |
| 8 | 0.29 | 0.34 | 0.26 | 0.30 | 0.23 | 0.28 | 0.96 | 0.97 | 3.07 | 3.23 | 1.68 |
| 9 | 0.30 | 0.36 | 0.27 | 0.33 | 0.25 | 0.29 | 0.92 | 0.97 | 3.07 | 3.25 | 1.69 |
| 10 | 0.31 | 0.37 | 0.27 | 0.37 | 0.27 | 0.31 | 0.90 | 0.97 | 3.01 | 3.29 | 1.70 |
| 11 | 0.32 | 0.38 | 0.28 | 0.40 | 0.29 | 0.33 | 0.88 | 0.96 | 3.07 | 3.31 | 1.72 |

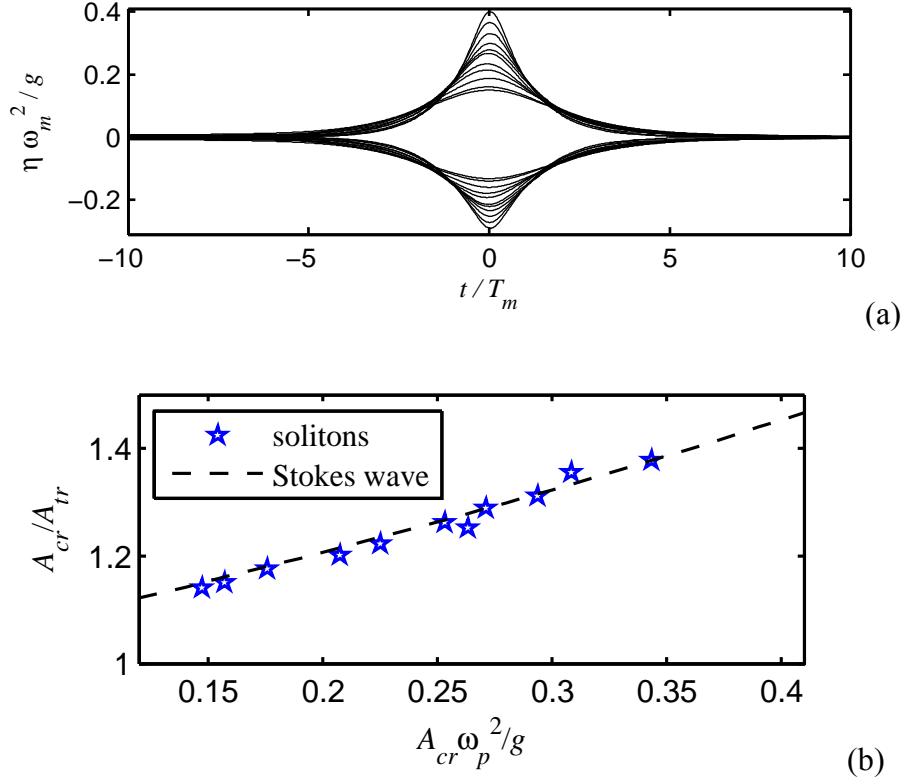

FIG. 3. Envelopes of the stationary wave groups which are observed in numerical simulations, for values $k_0 A_0$ = 0.15, 0.16, 0.20, 0.22, 0.23, 0.25, 0.28, 0.29, 0.30, 0.31, 0.32 (a), and the wave envelope vertical asymmetry $A_{cr} / A_{tr}$ as function of dimensionless crest amplitude (stars) (b). Dashed line in (b) shows the asymmetry of the uniform Stokes wave, when the strongly nonlinear correction to the frequency is taken into account.

The stationary wave group which is generated in the case $k_0 A_0$ = 0.30 is displayed in Figure 4 in more detail. Figures 4a,b show the wave envelope (dashed line) and the surface elevation (solid lines) at the moments when the maximum wave crest (thick solid line) and the deepest wave trough (thin solid line) occur, as functions of coordinate and time correspondingly. The different number of waves in the envelope in Figure 4a (space series) and Figure 4b (time series) is very notable and is the result of the difference between the phase and group velocities. The wave group in Figure 4a is steep, and consists of only about three waves, while the wave sequence in Figure 4b represents a visually smoother modulation of wave train.

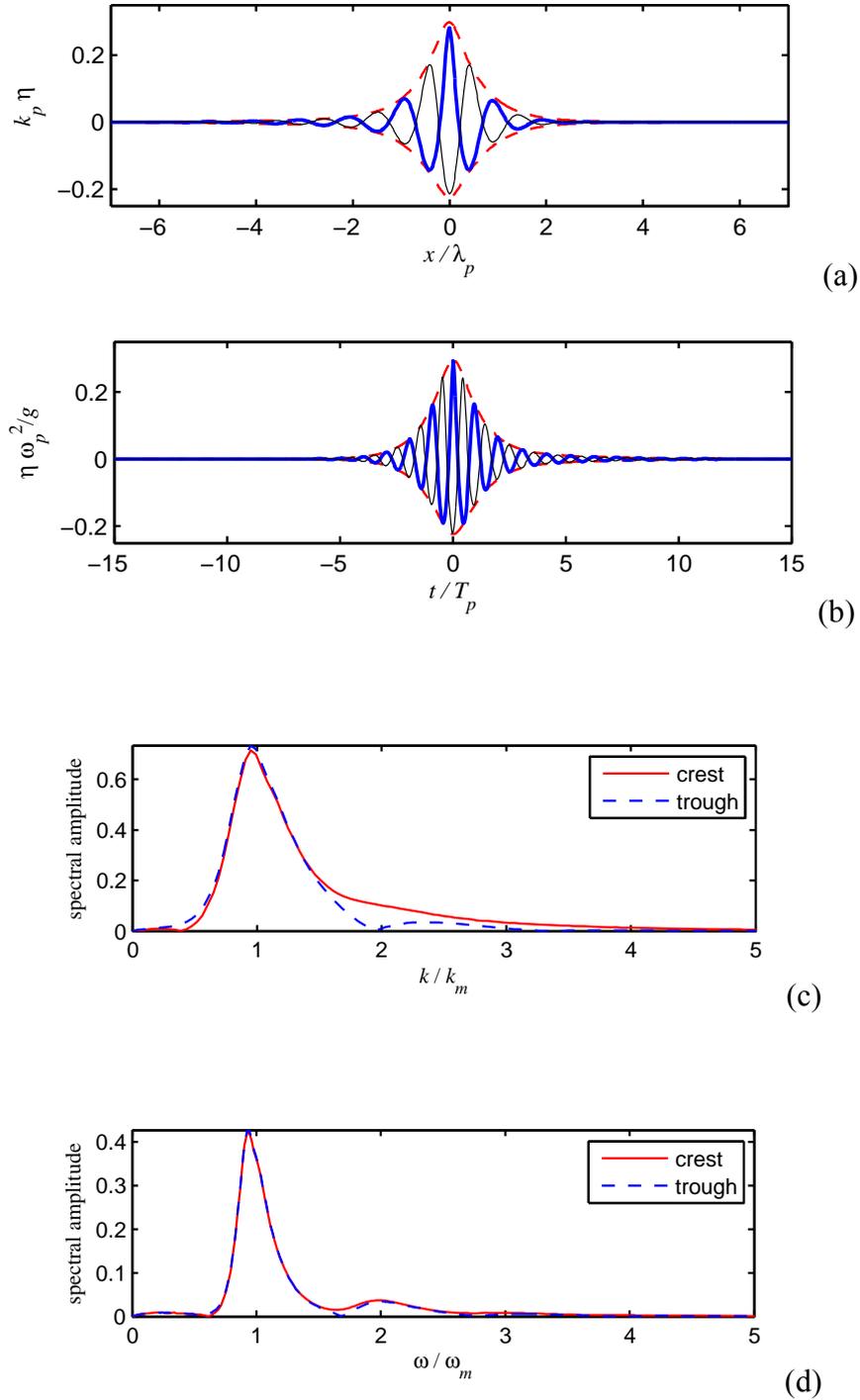

FIG. 4. The stationary wave group generated from the initial condition characterized by $k_0 A_0 = 0.30$ (experiment No 9 from Table I). Surface elevations (solid lines) and wave envelopes (dashed lines) are given in panels (a) and (b) as functions of coordinate and time

respectively. The wavenumber spectrum and frequency spectrum for the moments of maximum wave crest (solid lines) and the deepest trough (dashed lines) are shown in panels (c) and (d).

The wavenumber and frequency spectra of corresponding surface elevations are shown in Figure 4c and Figure 4d correspondingly. The spectra are computed for two phases of the wave group evolution: when the crest is highest (solid lines), and when the trough is deepest (dashed lines). While the NLS approximate solution possesses a stationary spectrum (see Section II and formula (5)), the evolution of spectrum is evident in Figures 4c,d, and is most notable in the case of wavenumber spectrum. In the moment of maximal wave crest the comb-shaped wavenumber spectrum becomes single-peaked. In the long-wave range the picture of spectrum evolution is inverse: the wavenumber spectrum has larger values at the moment of deepest through. In the frequency spectrum the second harmonic remains detached, and variation in the low-frequency range is negligible.

The spectrum evolution within a wave period becomes less visible for smaller waves. In the case $k_0 A_0 = 0.20$ the evolution cannot be seen by eye in the frequency spectrum, but may be noticed in the wavenumber spectrum plot. The persistence of frequency spectrum claims that the effect of spectrum evolution is most likely due to the overlap between the free and bound wave components, which occurs differently at different phases of the wave group evolution.

## IV. LABORATORY TESTS

Laboratory tests of steep and short solitary-like groups of waves were performed in the seakeeping basin of Technical University of Berlin. The basin is 110 m long, with a measuring range of 90 m. The width is 8 m and the water depth is 1 m. On the one side a fully computer controlled electrically driven wave generator is installed which can be utilized in piston as well as flap type mode. On the opposite side a wave damping slope is installed to suppress disturbing wave reflections. For this test campaign the wave generator is driven in flap type mode and the center of rotation is on the bottom of the basin as the wave board covers the full water depth.

The boundary condition for the wave maker, i.e. the wave sequence at the wave board, was obtained according to the following three different approaches:

1. in form of the NLS envelope soliton (4) for the free wave component; the surface elevation is $\eta = \mathrm{Re}(A(t)\exp(i\omega_0 t))$ (Method 1);

2. with use of time series of the surface elevation, obtained through strongly nonlinear simulations of stationary wave groups, as it is described in Section III (Method 2);

3. in form of the NLS envelope soliton (4), when the surface elevation $\eta(t)$ is reconstructed with help of formulas which take into account three asymptotic orders of the generalized NLS theory [17] (Method 3).

In different runs wave groups in two different phases (highest crest or deepest trough) were reproduced at the wavemaker.

Multiplication of the calculated wave sequence with the hydrodynamic as well as electric transfer function of the wave generator in frequency domain and subsequent Inverse Fast Fourier Transformation result in the control signal for the wave generator. The hydrodynamic transfer function is modelled using the Biesel function[22], relating the wave board stroke to the wave amplitude at the position of the wave maker. The obtained control signal is afterwards checked against the wave generator limitations – maximum wave board velocity and acceleration – to ensure a smooth operation as well as non-breaking waves at the wave board. The exact transfer of calculated boundary conditions to the seakeeping basin is a delicate procedure, in particular for this test campaign as deviations at the beginning will influence the complete propagation of the wave group along the basin. One point thereby is the application of linear transfer functions – for steeper waves the control signal may become more inexact in comparison to less steeper waves. But the main point is the fact that the wave generator provides a velocity profile at the wave board which differs significantly from the "natural" one under the same surface elevation depending on the geometry and type of the wave board. In this study the flap type mode of the wave board generates a velocity profile which decays linearly from top to bottom. The deviation between the flap motion and the exact particle motion under the surface elevation causes a first order disturbance at the wave board, which decays during propagation of the waves and is theoretically zero after a distance of only one wave length from the wave maker [23]. The influence of second order disturbances is significantly reduced due to the fact that the flap type mode is chosen and all generated wave lengths are almost in the deep water domain (all waves are deep-water waves regarding the criterion $h / \lambda > 0.5$, where $h$ is the water depth. The longest carrier wave length is about 1.8 m) [22]. So the waves need some time and space to become fully "physical" which maybe also influences the accuracy of the reproduced shape of the waves.

Figure 5 presents the general overview on the experimental setup. Altogether, ten wave gauges were installed: single devices at fetches 10, 30, 60 and 85 m, and an array of 6 closely situated gauges at the distance 45 m. The gauges will be numerated hereafter in the order how they appear from the wave maker to the other side of the tank. Special attention was paid on the position of wave gauge 1. As mentioned above, the selection of the first position follows two complementary requirements: on the one hand one is interested to be as close as possible at the wave board for comparison between target and reproduced wave; on the other hand the wave needs some time to fully evolve and thus the wave gauge should be placed several wave length away from the wave board. For this study the first wave gauge is placed 10 m in front of the wave maker which means that the wave gauge is at least 5 wave lengths away from the wave maker.

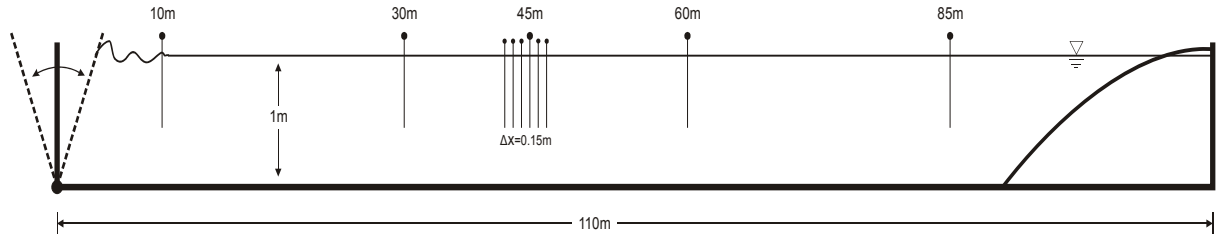

FIG. 5. Test setup – side view on the seakeeping basin with the wave generator on the left, the damping slope on the right as well as the positions of the ten wave gauges installed for this test campaign.

The test program comprises the variation of the initial wave steepness ($k_0A_0$ = 0.15, 0.2, 0.25, 0.3, 0.35) and the carrier wave frequency ($\omega_0$ =5.92, 6.82, 6.86, 7.52 rad/s, and corresponding $k_0h$ = 3.57, 4.74, 4.80, 5.76). The carrier frequencies are chosen in such a way that the relevant frequency bandwidth of the wave sequence spectrum is within the wave generator limitations. The lower reproducible frequency is 0.5 rad/s, and the highest frequency varies from 10 to 30 rad/s. Totally 43 runs with different initial conditions were measured. The focus was made on generation of wave groups which would not exhibit significant structural variation along the tank, and on excitation of maximum steep steady wave groups. Only conditions without evidence of wave breaking were permitted. Parameters for the best experiments, when quasi stationary wave groups were observed, and for one case of unstable wave packet, are given in Table II.

Many of performed experimental runs exhibit significant destruction of the wave group as it propagates along the tank. Such an example is given in Figure 6a, which shows results of the measurements; they are ten time series of the surface elevation registered by gauges from 1 to 10. A radiated wave train behind the main wave group is well seen in Figure 6a at gauges 1 and 2 (right side from the main group). Later on, a rather intense wave train outruns the main wave group at gauges 2-10. The initial condition in this case is generated according to Method 3.

Table II. Initial conditions for selected laboratory experiments.

| Experiment code | $k_0A_0$ | $\omega_0$, rad/s | $k_0h$ | Maximum frequency reproduced by the wave maker, $\omega_{max}$, rad/s | Method for wavemaker signal | Wave phase | Structural stability |
|---|---|---|---|---|---|---|---|
| 29.14 | 0.2 | 6.86 | 4.80 | 12 | 2 | crest | stable |
| 30.07 | 0.2 | 6.86 | 4.80 | 12 | 2 | trough | stable |
| 30.13 | 0.3 | 6.86 | 4.80 | 12 | 2 | crest | stable |
| 30.16 | 0.3 | 6.86 | 4.80 | 12 | 2 | trough | stable |
| 30.29 | 0.3 | 5.92 | 3.57 | 20 | 1 | crest | unstable |
| 30.37 | 0.35 | 6.82 | 4.74 | 15 | 1 | crest | stable |

Figure 6b shows measurements of a stable wave group which is observed in experiment No 30.16. The initial signal is the time series taken from the numerical simulation of a stationary wave group (Method 2). Both experiments shown in Figure 6 correspond to the same initial wave group steepness, $k_0A_0 = 0.3$.

Sometimes measurements from gauge 10 contained spikes; a short frequency cut-off filter was applied in this situation.

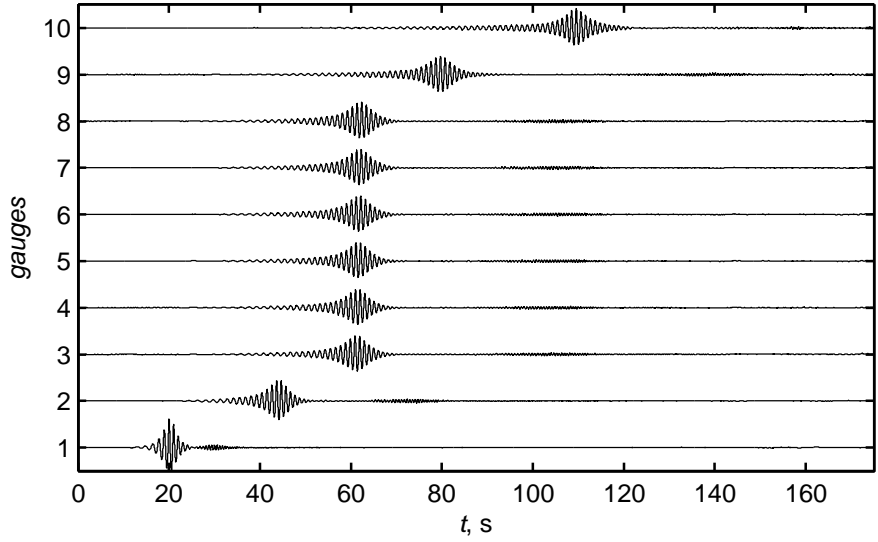

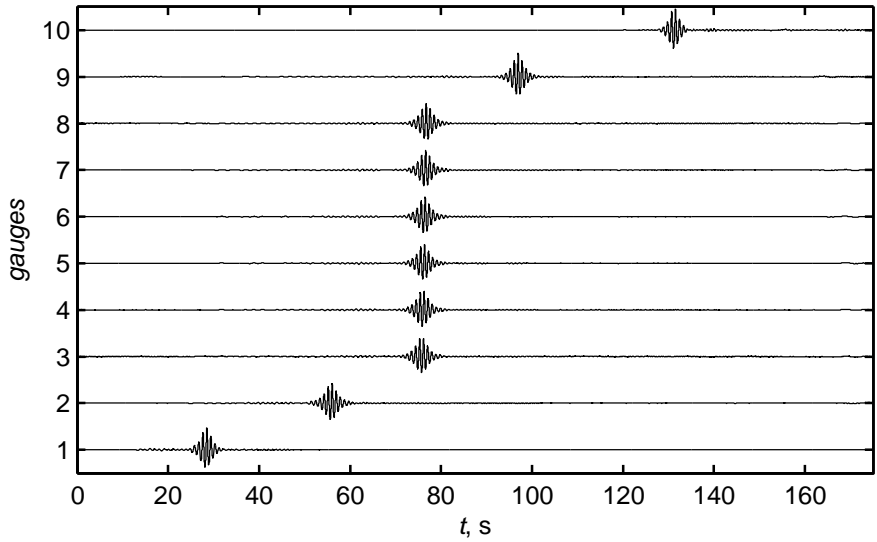

FIG. 6. Time series of the surface elevation at different distances, measured in the laboratory tank: an unstable wave group (a) (experiment No 30.29), and stationary wave group (b) (experiment No 30.16). Both the cases correspond to $k_0 A_0 = 0.3$, but to different carrier wave frequencies and different methods of signal generation.

For each experimental run, the velocity of the wave group, $V$, was obtained tracking its path, similar to the approach applied previously to the numerical experiments. Then all the records from ten gauges were plotted in the moving with velocity $V$ reference system. The changeability of the envelope in the course of the wave group propagation was estimated visually, and cases with most permanent wave group shapes were picked out as the 'best' ones (they are listed in Table II) and studied further. The surface elevation profiles for the best experiments are shown by thin solid lines in Figures 7a-c in co-moving coordinates. One of the time series, at gauge 3, is given by a thicker line.

For the case of initially relatively small-amplitude waves, $k_0A_0 = 0.20$, experiments No 29.14 and No 30.07 are found to be the best (Figure 7a); for $k_0A_0 = 0.3$ – experiments No 30.13 and No 30.16 (Figure 7b, time series of the latter experiment are shown in Figure 6b); for $k_0A_0 = 0.35$ – experiment No 30.37 (Figure 7c). Parameters of initial conditions for the experiments are given in Table II.

Enveloping dashed curves in Figures 7a-c come from the strongly nonlinear numerical simulations described in the previous Section. Note that the best fit between the envelopes and the displacement shapes is obtained when the steepness of the initial condition, $k_0A_0$, is different for the cases of laboratory experiments and the numerical simulations. The experimental cases shown in Figure 6a ($k_0A_0 = 0.20$), Figure 6b ($k_0A_0 = 0.30$) and Figure 6c ($k_0A_0 = 0.35$) are compared to the numerical simulations of groups $k_0A_0 = 0.15$, $k_0A_0 = 0.23$ and $k_0A_0 = 0.29$ respectively. Thus, for the same initial conditions, stationary wave groups seem to have eventually smaller steepness in laboratory experiments in comparison to the numerical simulations. The wave steepness obtained in the physical basin is smaller most likely due to the above mentioned peculiarities of the boundary conditions of the wave generator or due to the wave dissipation near the wavemaker. For the three considered cases the steepnesses of stationary wave groups are $A_{cr}\omega^2_m/g \approx 0.150, 0.235, 0.301$ respectively, see horizontal dotted lines in Figure 7. The quantity $\omega^2_m/g$ gives the estimation of the mean wavenumber, which is directly calculated on the basis of the measured time series.

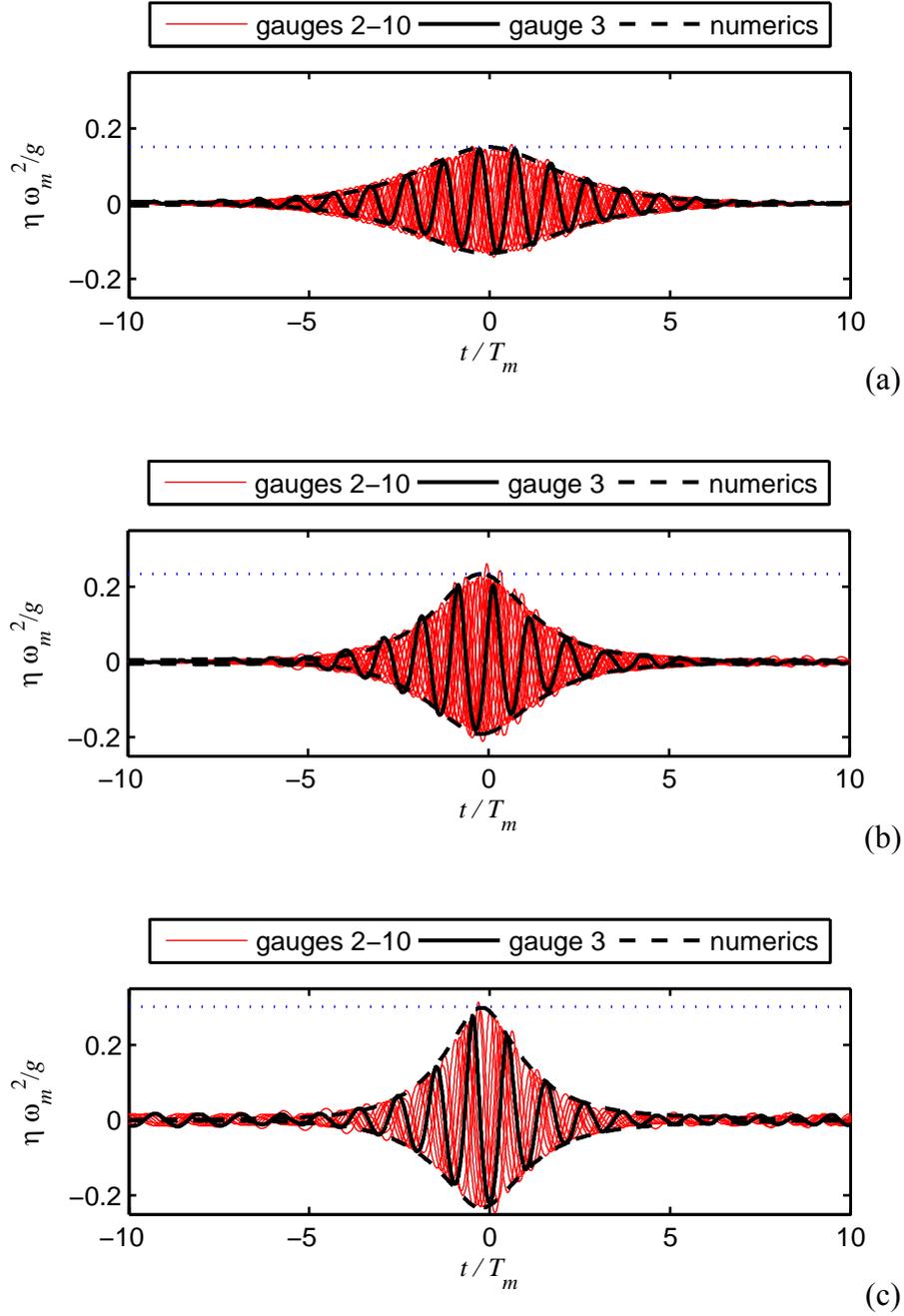

FIG. 7. Stationary wave groups: comparison between laboratory and numerical results. The sequence of thin solid lines is the time series of surface elevations registered by 10 gauges, when plotted in co-moving references. The panels show results of experiments 29.14 and 30.07, $k_0 A_0 = 0.20$ (a), experiments 30.13 and 30.16, $k_0 A_0 = 0.30$ (b) and experiment 30.37, $k_0 A_0 = 0.35$ (c). The series from gauge 3 is given by a thicker line. The enveloping curves (dashed lines) are obtained in the strongly numerical simulations of the Euler equations with appropriate amplitudes of the initial condition: $k_0 A_0 = 0.15$ (a) $k_0 A_0 = 0.23$ (b) and $k_0 A_0 = 0.29$

(c) (simulations 1, 5 and 8 from Table I respectively). Horizontal dotted lines mark the scaled envelope crest amplitudes, which are $A_{cr}\, \omega_m^2/g \approx 0.150, 0.235, 0.301$ for cases (a), (b) and (c) respectively.

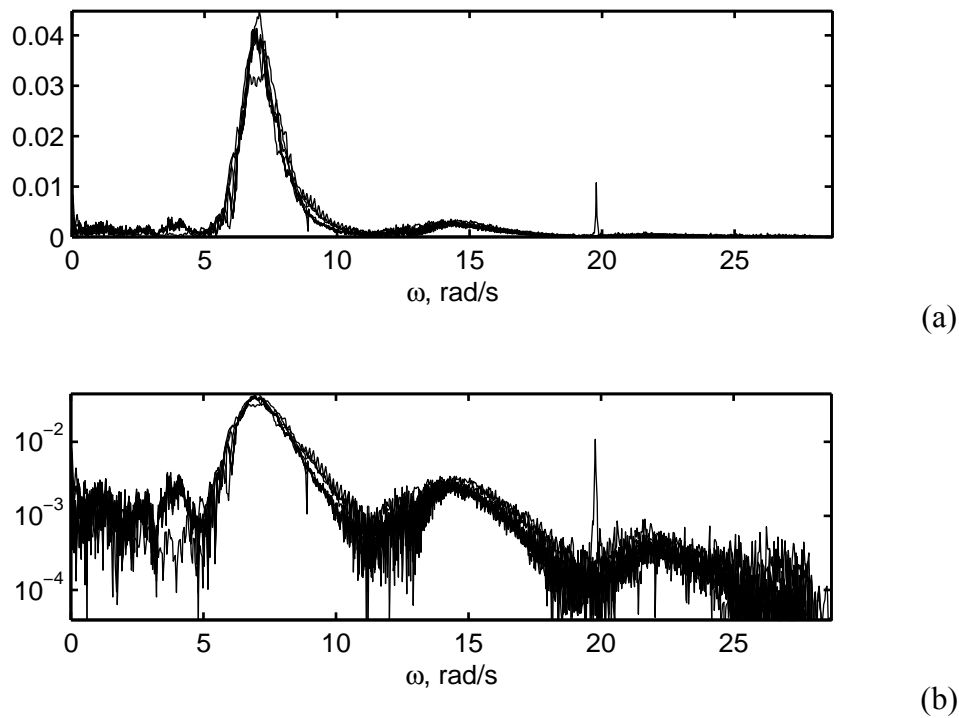

FIG. 8. Scaled frequency Fourier amplitude spectrum for experiment No 30.16 in linear (a) and semi-logarithmic (b) coordinates. The lines show results from all 10 gauges.

A very good agreement between the envelope shape obtained in the numerical simulations and the sequences of wave elevation profiles from the laboratory experiments (using the adjusted wave steepness of the initial condition) may be pointed out. Figure 7c shows the most extreme non-breaking stationary wave group which was observed in the laboratory conditions. It fits well the almost most extreme stationary wave group found in the numerical simulations.

The frequency spectrum for laboratory experiment No 30.16 ($k_0 A_0 = 0.30$) is shown in Figure 8 in linear and semi-logarithmic scales; ten records from the gauges are given in one plot. Generally, two wave harmonics may be clearly distinguished in the spectra of stable wave groups, and the third harmonic is somewhat less evident, see Figure 8b. The spectrum in semi-logarithmic scales looks rather stationary, while in the linear scales some variations may be observed. Such variability seems to be much less pronounced in experiments with less steep wave groups (No 29.14 and 30.07) in comparison with other steeper cases listed in Table II. Numerical simulations do not report such strong variability of the Fourier spectrum, see Figures 4c,d.

Velocities of the stationary wave groups observed in laboratory experiments are given by circles in Figure 9 for the best selected cases of stable wave groups. The velocities are scaled with the linear wave group velocity, estimated for time series as $C_{gr} = g/(2\omega_m)$. There are five markers for three values of dimensionless upper amplitudes of the wave groups, $A_{cr}\omega_m^2/g$. By stars the same characteristics for numerical simulations of hydrodynamic equations are plotted. Perfect agreement between the laboratory and numerical results is found.

Two pairs of the group amplitudes almost coincide in Figure 9. These pairs correspond to the boundary conditions with the same steepness ($k_0A_0 = 0.20$ and $k_0A_0 = 0.30$ respectively) but with different phase of the wave group: 'crest' or 'trough' (see Table II). Hence, the resulting stable wave group is proved to be independent on the phase of the wavemaker signal; simultaneously, the laboratory measurements are shown to be repeatable.

The propagation speed of the NLS soliton (4) is equal to the linear wave group velocity and does not depend on the wave amplitude. Velocities of stationary wave groups observed in the laboratory and numerical experiments obviously depend on the amplitude as shown in Figure 9. The velocity grows with amplitude; it exceeds about 11% the value of the group velocity of linear waves $C_{gr}$ in the steepest laboratory case, and about 16% – in the numerical simulations.

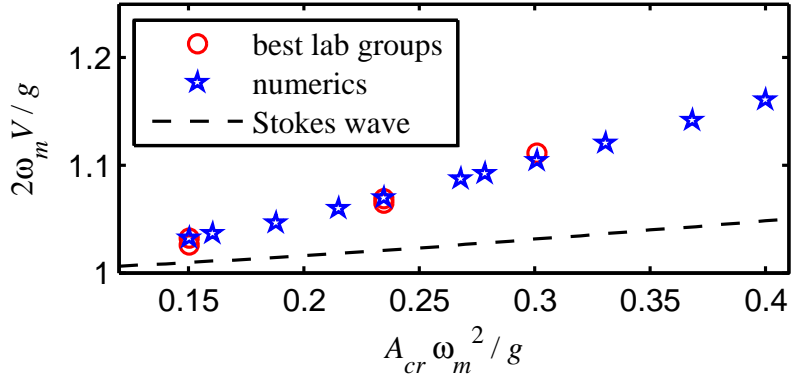

FIG. 9. Velocities of stationary wave groups, observed in numerical simulations (stars), and of the selected 'best' wave groups measured in laboratory experiments (circles). The speed of the uniform Stokes wave is given by the dashed line for the reference.

The dashed line in Figure 9 shows the nonlinear velocity of Stokes waves (which is found numerically) normalized by the linear speed $C_{gr}$. It is clear that the effect of nonlinearity on the speed of solitary waves is even more significant than the nonlinear correction to the Stokes wave velocity.

The general conclusion which may be made on the basis of all 43 experimental runs is that it is difficult to decide, which approach between Method 1 and Method 2 is the best. Both of these methods provide significantly better initial conditions than Method 3. As waves with $k_0A_0 = 0.35$ break in numerical simulations, for this steepness only Method 1 is able to provide an initial condition. We emphasize, that Method 1 – is the basic analytic formula for the exact envelope soliton solution of the NLS equation, when bound waves are disregarded. A difference between results of experiments, when the boundary condition is represented by a group with high crest or with a deep trough, is not found. In the steepest case $k_0A_0 = 0.35$ experiment No 30.37 starts from the wave group with a high crest (see Table II). An experiment with the same steepness $k_0A_0 = 0.35$ of the boundary condition, but when the group has a deep trough instead of a high crest, was not conducted.

The particular choice of the carrier frequency is seemingly important for the successful generation of a steep solitary group; the reason for that is, however, not fully identified. The experimental conditions correspond to not very deep water; for longer carrier wave the effect of finite depth may become important (this seems to be the only substantial difference between experiments No 30.29 and No 30.37 in Table II). Simultaneously, variation of the mean wave frequency changes the relative limits of the frequency domain, which can be reproduced by the wave generator. Hence, this circumstance also may have effect on the wave generation.

## V. CONCLUSIONS

The existence of structurally stable short wave packets of steep waves is proved in laboratory experiments; they propagate without noticeable change of envelope shape for more than 60 wavelengths, what is about 15-30 lengths of the group depending on how the length is defined. To the best of our knowledge, so steep solitary wave groups (up to $A_{cr}\omega^2_m/g \approx 0.30$, where $\omega_m$ is the mean frequency and $A_{cr}$ is the maximum wave crest amplitude) are observed in water wave tank for the first time. Similar solitary groups of intense waves were found recently in numerical simulations of the primitive Euler equations [13,14].

The laboratory measurements are compared with the results of strongly nonlinear numerical simulations of the Euler equations, where stationary wave groups are obtained as a result of long-time evolution of NLS envelope solitons. Enveloping curves for the numerically obtained stationary wave groups fit surface elevations of stable groups observed in the laboratory experiments very well. Velocities of the stable wave groups, obtained on the basis of laboratory and numerical experiments, agree perfectly. The numerical simulations reveal spectral variation of stationary wave groups, most likely due to the overlap between free and bound wave harmonics. Laboratory measurements claim some stronger spectral variation for the most intense wave groups; its genesis is not clear.

When compared with uniform strongly nonlinear Stokes waves with the same frequency and crest amplitude, the observed solitary groups do not show distinctive difference in vertical asymmetry. However, the solitary groups move significantly faster than the linear waves; the effect of nonlinearity on their speed is even stronger than the nonlinear correction to the Stokes wave velocity.

Generally speaking, the NLS analytic solution for free waves is found to be quite efficient for generation of fairly steep solitary wave groups in laboratory conditions. The observed groups correspond to the limiting case (very short and very intense) of envelope solitons, known for the weakly nonlinear NLS framework. Results of laboratory experiments turn out to be sensitive with respect to the change of carrier wave frequency. Most likely this is due to the change of dimensionless water depth, $k_0 h$.

It seems so that this is the wave breaking effect, which eventually limits stationary wave groups in amplitude. Even steeper and shorter long-living wave packets are observed in the 2D numerical simulations (up to $A_{cr}\, \omega_m^2 /g \approx 0.40$) in comparison with the laboratory observations ($A_{cr}\, \omega_m^2 /g \approx 0.30$). These maximal groups are formed from NLS envelope solitons with steepness $A_0 \omega_0^2 /g = 0.32$ and $A_0 \omega_0^2 /g = 0.35$ respectively.

In laboratory experiments the efficiency of the initial (boundary) condition seems to be worse, although no obvious wave radiation, which could result in a wave group damping, was found in the records of the "best" stable wave groups. For steeper waves the control signal for the wave maker via linear Response Amplitude Operators may become more inexact in comparison to less steep waves. Another important reason for the loss of efficiency of generation of steep solitary groups could be the geometry and type of wave board, i.e. depending on the wave board mode and geometry the velocity profile at the wave board differs significantly from the "natural" one under the same surface elevation. It should be stressed however, that other physical mechanisms could limit the maximum height of the quasi-unidirectional stationary group observed in the laboratory tank, such as 3D instability effects[1]. Indeed, intense waves are known to suffer from 3D instabilities; however, the essentially strong modulation is the peculiarity of the considered waves and hence the direct adoption of results of stability analysis for uniform or weakly modulated waves is not possible. Conditions which led to wave breaking were not considered in the present work.

**ACKNOWLEDGMENTS**

The research has received funding from the EC's Seventh Framework Programme FP7-SST-2008-RTD-1 project EXTREME SEAS - Design for Ship Safety in Extreme Seas (http://www.mar.ist.utl.pt/extremeseas/) under grant agreement No 234175. AS acknowledges partial support from the EC's Seventh Framework Programme FP7-PEOPLE-2009-IIF under grant agreement No 254389/909389, and also RFBR 11-02-00483 and 12-05-33087. M.O. has been funded by ONR grant N000141010991.

AS is grateful to colleagues in the Department of Mathematics and EPSAM at Keele for the warm hospitality.

---

[1] Thanks to the anonymous referee who suggested this possible explanation.